# Atomic-Void van der Waals Channel Waveguides

Haonan Ling[1], Jacob B. Khurgin[2], Artur R. Davoyan[1*]

1. Department of Mechanical and Aerospace Engineering, University of California, Los Angeles, California 90095, United States
2. Department of Electrical and Computer Engineering, Johns Hopkins University, Baltimore, Maryland 21218, United States

*Corresponding author: davoyan@ucla.edu

**Abstract**

Layered van der Waals materials offer a unique platform for creating atomic-void channels with sub-nanometer dimensions. Coupling light into these channels may further advance sensing, quantum information, and single molecule chemistries. Here we examine limits of light guiding in atomic-void channels and show that van der Waals materials exhibiting strong resonances – excitonic and polaritonic – are ideally suited for deeply subwavelength light guiding. We demonstrate that excitonic transition metal dichalcogenides can squeeze >70% of optical power in just $< \lambda/100$ thick channel in the visible and near-infrared. We also show that polariton resonances of hexagonal boron nitride allow deeply subwavelength ($< \lambda/500$) guiding in the mid-infrared. We further reveal effects of natural material anisotropy and discuss the influence of losses. Our analysis shows van der Waals channel waveguides while offering extreme optical confinement exhibit significantly lower loss compared to plasmonic counterparts. Such atomic void waveguides pave the way to low loss and deeply subwavelength optics.

**Introduction**

Layered van der Waals (VdW) materials have emerged as a promising material platform with unique optical, electronic and mechanical properties [1-7]. Weak interlayer bonding allows isolating individual layers with naturally passivated atomically flat surfaces. Such control of material thickness with ångström precision paves the way to creating diverse structures with atomic-scale features [8-10]. An interesting class of such systems are *atomic-void* geometries in which one or several atomic layers are judiciously removed to create an atomically thin channel [11-14]. Studies of such atomic-void channels made of different VdW materials, including graphene, hexagonal boron nitride (hBN), and transition metal dichalcogenides (TMDCs), have revealed intriguing transport physics. Such effects as ballistic molecular transport [12], field effect switching of ionic transport [14], and apprising elusive interfacial phenomena

[13] are only a few of the demonstrated ones. Coupling light into atomic-void channels may enable precise control of physical [15-18] and chemical processes [19] at the level of individual molecules, as well as high precision quantum sensing and readout [20,21]. However, limits of light squeezing and guiding in atomic-void channels and cavities remain elusive, particularly, in channels that are orders of magnitude smaller than the free-space wavelength of light ($\ll \lambda/100$, $\lambda$ is the free-space wavelength).

Light confinement and guiding well beyond the diffraction limit (i.e., $\ll \lambda/2$,) has been an active area of study [22,23] that is critical to a number of emergent applications, including quantum control [21,24], near-field imaging [25,26] and sensing [27-31] optical manipulation [32-34], and high-speed interconnects [35-37]. Subwavelength light control is important also for stimulating strong light-materials interaction to achieve higher nonlinear responses and controlling quantum phenomena at the nanoscale [24,38,39]. Deeply subwavelength guiding and manipulation of optical fields is possible with plasmonic nanostructures [22,23,40,41]. However, their inherently lossy nature [42,43] limits the scope of possible applications. On the other hand, use of all-dielectric structures can mitigate the undesired absorption thus enabling low-loss systems [43,44]. One example are dielectric slot waveguides, where light is confined in a channel sandwiched between high refractive index slabs (Fig. 1a) [45,46]. For instance, recently demonstrated Si and hBN slot waveguides are capable of efficiently guiding near-infrared and visible light, respectively, within just 50-80 nm wide channels (i.e., $\sim 0.06\lambda$ for Si at $1550\,nm$ and $\sim 0.15\lambda$ for hBN at $534\,nm$) [46,47]. Such all-dielectric subwavelength waveguides are of great interest for biosensing [29,30], electro-optical modulation [48-50], and quantum light manipulation [47]. However, unlike plasmonic nanostructures, relatively small refractive indexes of Si and hBN in near-infrared and visible bands set a limit on the width of the channel in which light can be efficiently squeezed and guided. Guiding light efficiently in nanometer ($\leq \lambda/100$) and in atomically thin ($\leq \lambda/1000$) channels requires a different material platform with a high index contrast

VdW materials exhibiting unusually high refractive indexes across different wavelength bands [51-53] and enabling atomically smooth surfaces with ångström precision are uniquely positioned for atomic-void light waveguides. Indeed, TMDCs slightly below their excitonic resonance exhibit refractive index that is higher than that of conventional covalent semiconductors (e.g., Si and GaAs) in the visible and near-infrared frequency bands [6,54-56]. hBN, on the other hand, possesses very sharp phonon-polariton resonances in the mid-infrared range resulting in a very large dielectric permittivity [57-60]. The greater part of VdW materials studies have been focused on single or a few layers' structures. At the same time, highly anisotropic bulk VdW materials also possess exceptional properties, high refractive index being one of them. While a number of diverse photonic devices made of these materials has been reported [55,56,61-63], the limits of optical confinement and guiding in atomically small gaps [11-14] remain unexplored. Here, inspired by the slot waveguide configuration, we examine theoretically atomic-void waveguides made of TMDCs and hBN VdW materials. We demonstrate that efficient squeezing and low-loss light guiding within nanometer and sub-nanometer channels is possible. We further reveal effects of natural anisotropy of bulk VdW materials on lightwave confinement.

**Results and Discussion**

To understand how light can be guided in deeply subwavelength channels, we begin our discussion with a parametric study of a one dimensional atomic-void waveguide schematically shown in Fig. 1a. The waveguide is made of a thin channel of thickness $d$ created within high refractive index VdW slabs each with a thickness $w$. Such configuration can guide light in a cut-off free fundamental transverse magnetic (TM) mode (i.e., with the electric field $\boldsymbol{E} = (E_x, 0, E_z)$). The optical properties of this mode are governed by the continuity of the displacement field at the channel interface: $\epsilon_{channel} E_x^- \left(\frac{d}{2}\right) = \epsilon_{xx} E_x^+ \left(\frac{d}{2}\right)$, where $\epsilon_{channel}$ is the permittivity of the channel ($\epsilon_{channel} = 1$ for vacuum) and $\epsilon_{xx}$ is the component of the permittivity tensor perpendicular to the channel. Evidently as $\frac{\epsilon_{channel}}{\epsilon_{xx}} \to 0$ the electric field in the channel grows, resulting, as we show below, in a stronger confinement and smaller waveguide footprint. A related dispersion equation of the fundamental TM mode can be expressed as follows [45]:

$$\tanh\left(\frac{k_a d}{2}\right) = \frac{k_s}{k_a n^2} \frac{\sin(k_s w) - \frac{k_a n^2}{k_s} \cos(k_s w)}{\cos(k_s w) + \frac{k_a n^2}{k_s} \sin(k_s w)} \quad (1)$$

where $n = \sqrt{\epsilon_{xx}}$ is the refractive index of the slabs, $k_a = \frac{2\pi}{\lambda}\sqrt{\beta^2 - 1}$ and $k_s = \frac{2\pi}{\lambda}\sqrt{n^2 - \beta^2}$ with $\beta$ being the effective mode index. We note that in sub-nanometer channels nonlocal effects and effects of quantum tunneling may play a prominent role. However, classical approach provides a good first order estimate for assessing lightwave dispersion. Here we make use of a classical local model.

Dispersion and optical properties of such a channel waveguide depend strongly on the channel width, $d$, slab thickness, $w$, and the slab refractive index, $n$. In Fig. 1b we study the role of refractive index by examining the lightwave dispersion with the variation of the channel width, $d$, for a waveguide with slab thickness $w = 0.1\lambda$. As expected, light can be guided without cut-off at all channel widths. Waveguides with a higher refractive index exhibit higher effective index $\beta$. As the channel width decreases the effective index grows and in the limit of $d \to 0$ reaches $\beta \to n$. The latter, as we will show below, signifies strong light confinement within the channel.

Moving forward we examine optical power carrying capacity of the channel, which we define as: $C_p = \frac{\int_{-d/2}^{d/2} Re(\boldsymbol{E} \times \boldsymbol{H}^*) dx}{\int_{-\infty}^{\infty} Re(\boldsymbol{E} \times \boldsymbol{H}^*) dx}$, where $\boldsymbol{E}$ and $\boldsymbol{H}$ are electric and magnetic field vectors. Such definition measures the fraction of the optical power (Poynting vector) guided within the channel. Hence, $C_p \ll 1$ would imply that light is predominantly guided outside of the channel (e.g., in the cladding slabs), whereas $C_p \simeq 1$ would manifest that almost all the power is guided within the channel. Fig. 1c shows variation of factor $C_p$ as a function of refractive index, $n$, and slab width, $w$, for a channel with $d \leq 0.01\lambda$. Evidently, the higher is the refractive index the higher is power carrying capacity, $C_p$. Furthermore, for each $n$ value there exists an optimal slab width at which power carrying capacity for a given channel thickness is maximized. For a very small slab width, $w$, the field gets delocalized leading to a drop in power carrying capacity. However, with the increase of value of $n$ the optimal power carrying capacity is attained for thinner slabs. To further explore the limit of the electric field enhancement with the variation of the channel width and slab refractive index, in Fig. 1d we study electric field enhancement factor: $C_f = \frac{max_{channel}(|\boldsymbol{E}|^2)}{<|\boldsymbol{E}|^2>_{total}}$, where $<|\boldsymbol{E}|^2>_{total} = \frac{\int_{-\infty}^{\infty} |\boldsymbol{E}|^2 dx}{A_{total}}$, $max_{channel}(|\boldsymbol{E}|^2)$ represents the maximum value of electric field intensity within the channel and $A_{total}$ is the total cross-section of the waveguide. Interestingly, as the channel width decreases the electric field gets strongly enhanced. Therefore, one may expect that there is an optimal waveguide configuration in which maximum power and high electric field are attained simultaneously in the channel. We anticipate that as the index contrast at the channel interface grows, the optimal regime is observed in very narrow channels. Worth noting that for higher index contrast the effective mode area and related device footprint are dramatically reduced. In Fig. 1e we plot variation of the optimal effective mode width, $w_{eff}$, as a function of the refractive index, $n$, for several different channel widths (here the effective width, $w_{eff}$, measures $1/e$ optical field intensity drop outside the waveguide, see also Fig. 1e inset). As expected, the larger the value of $n$ is the smaller the minimum effective width and related device footprint.

The effects discussed here pertain to different frequency domains, including visible, near- and mid-infrared as long as materials with high refractive index are used to create a channel waveguide. VdW materials are particularly interesting in this regard owing to their atomic thickness and resonant material responses. Below we study specific examples of waveguides made of excitonic TMDCs and polaritonic hBN.

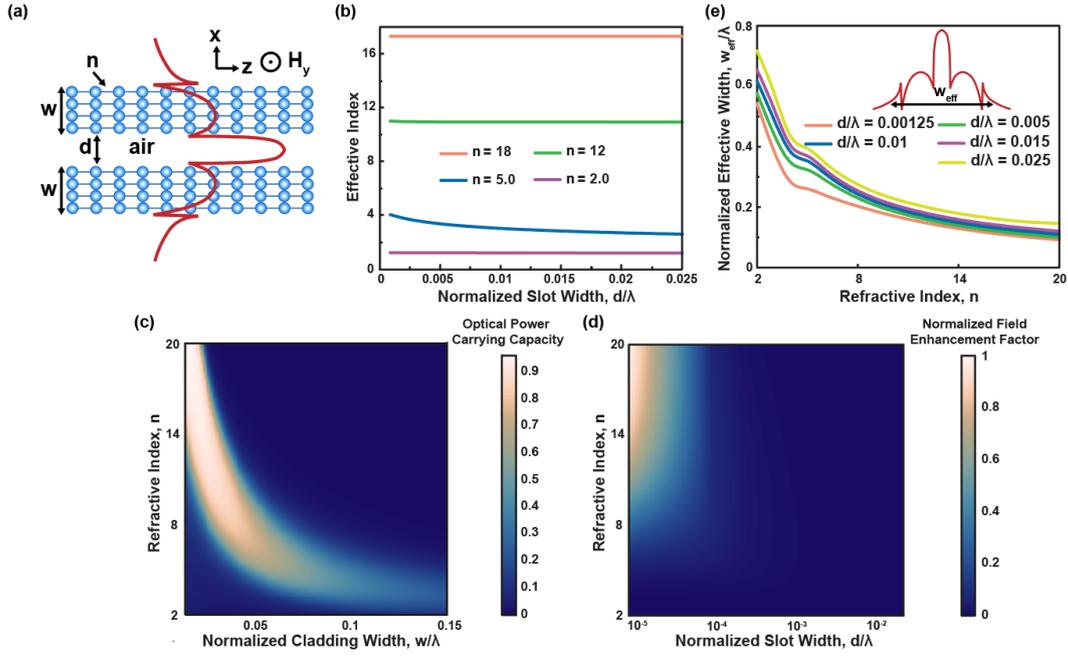

FIG. 1. One dimensional atomic-void VdW channel waveguide. (a) Schematic illustration of a waveguide. Typical electric field profile is shown. (b) Effective index dispersion with slot width for different slab refractive indexes. Here $w = 0.1\lambda$. (c) Power carrying capacity of a channel waveguide as a function of slab width, w, and refractive index, $n$. Here $d = 0.01\lambda$ is assumed. (d) Normalized electric field enhancement factor in a channel waveguide as a function of channel width, d, and slab refractive index, $n$. Here $w = 0.1\lambda$. (e) Normalized effective waveguide width, $w_{eff}/\lambda$, with variation of slab refractive index for different slot widths.

TMDCs are an emerging class of layered VdW semiconductors [2,6,7]. Many materials in this family, including MoS₂, WS₂, MoSe₂, and WSe₂, exhibit high refractive index [6,7,54-56]. We identify three fundamental factors that determine high refractive index of bulk layered TMDC materials Fig. 2a inset). First of all, permittivity is a measure of optical polarizability, i.e., the ability of valence electrons to oscillate between the ions. However, the ability to freely oscillate between the ions also means strong coupling between the states, manifested by broadening of the conduction and valence bands. As a result, the peak of the joint density of states and thus absorption are situated far above the absorption edge and according to simple Kramers-Kronig considerations the permittivity (and related refractive index) in the transparency range is limited. This restriction, however, is partially mitigated in layered VdW materials in which the relatively weak interlayer bonds do not cause much broadening (Fig. 2a inset) [64-66]. As a consequence, the in-plane permittivity can be relatively high at the expense of low out-of-plane permittivity. For most applications, including the channel waveguides studied here, high index for light polarized in just one or two directions is sufficient.

The second reason for high index of TMDCs is the d-shell character of the valence and conduction bands which both provides large number of valence electrons and makes bands relatively flat, thus bringing peak of the joint density of states closer to the absorption edge (Fig. 2a inset) [64-66].

Finally, even in multilayer TMDCs strong excitonic resonance can be observed. The fact that multilayer TMDC is indirect gap material is irrelevant here – as long as strong excitonic peaks are located just slightly above the absorption edge and contribute to increased permittivity via Kramers Kronig relations.

Moving forward, for the sake of concept demonstration we study $MoS_2$ channel waveguides. Other TMDC materials are expected to exhibit a similar performance. $MoS_2$ possesses a strong anisotropy which is manifested in a large difference between in-plane and out-of-plane refractive index components (Fig. 2a) [67]. Therefore, two waveguide configurations are possible: One with a channel oriented along the crystal axis (Fig. 2b inset), and another one with a channel oriented along the crystallographic plane (Fig. 3a). We shall denote these two possibilities as "out-of-plane channel" and "in-plane channel" configurations, respectively. As the electric field is polarized along the *x*-direction, the two configurations will exhibit drastically different properties driven by in-plane and out-of-plane material responses (Fig. 2a). For the sake of completeness below we study both configurations.

We begin our study with an out-of-plane channel configuration (Fig. 2b inset). In this case wave dispersion is governed predominantly by the in-plane component of the refractive index, which exhibits a strong excitonic resonance (Fig. 2b). As a result, such configuration is expected to demonstrate the highest power carrying capacity and field enhancement. At the same time, fabrication of such ultranarrow channels might be rather challenging. It is worth noting recent progress in lateral growth [68] and anisotropic etching [69].

First, we examine the dispersion of power carrying capacity in the channel for two different waveguide dimensions, as shown in Fig. 2b. By tuning the waveguide dimensions optimal power carrying capacity can be attained in a desired wavelength range. Hence, for a structure with a narrower channel (i.e., $d = 10\ nm$) the maximum power carrying capacity is attained at wavelength $\lambda \sim 680\ nm$, i.e., at the $MoS_2$ excitonic resonance [2,6,7]. A wider channel (i.e., $d = 35\ nm$), on the other hand, guides power more efficiently in the sub-bandgap regime ($\lambda \sim 1200\ nm$). In both cases we observe that $\geq 70\%$ of power can be guided within a very narrow channel (compare with a ~30% in an equivalent silicon slot waveguide) [45].

Next we examine power carrying capacity and electric field enhancement with the variation of the waveguide geometry at two respective characteristic wavelengths: $\lambda = 680\ nm$ and $\lambda = 1200\ nm$. In Figs 2c and 2e, we plot power carrying capacity, $C_p$, as a function of the slot width for several different dimensions $w$ and $h$ at $\lambda = 680\,nm$ and $\lambda = 1200\,nm$, respectively. At the excitonic wavelength, $\lambda = 680\,nm$, high refractive index of bulk MoS$_2$ results in a very efficient confinement of optical power within a deeply subwavelength channel, as shown in Fig. 2c. Indeed, for $w = 40\,nm$ and $h = 260\,nm$, over 77% of the power can be transmitted in a just 18 nm-wide channel, which is ~40 times smaller than the free-space wavelength of light. Furthermore, even nanometer and sub-nanometer channels (e.g., $d = 0.7\,nm$) exhibit very efficient power carrying capacity, i.e. $C_p \simeq 35\%$. In the more practical sub-bandgap regime (i.e., $\lambda = 1200\,nm$, shown in Fig. 2e) longer wavelength and lower index contrast at interface result in optimal waveguides with a wider channel in which smaller power carrying capacity is attained. However, even in this case over 70% of power can be squeezed in just a 50 nm-wide channel, and over 10% of the power when $d = 1\,nm$. Figs 2d and 2f show respective field enhancement factors, $C_f$, at $\lambda = 680\,nm$ and $\lambda = 1200\,nm$. As in the case of a one dimensional waveguide studied earlier, as the channel gets smaller the electric field inside it grows dramatically. Respective eigen-mode profiles (electric fields) for a $d = 5\,nm$ channel are plotted in the insets of Figs 2d and 2f. It is clearly seen that the optical field is strongly localized within the channel. For ultranarrow channels, such as a single atomic layer void ($d = 0.7\,nm$), at $\lambda = 680\,nm$ the optical field is enhanced by over $10^4$ (although fabrication of an equivalent atomically thin out-of-plane channels is not straightforward).

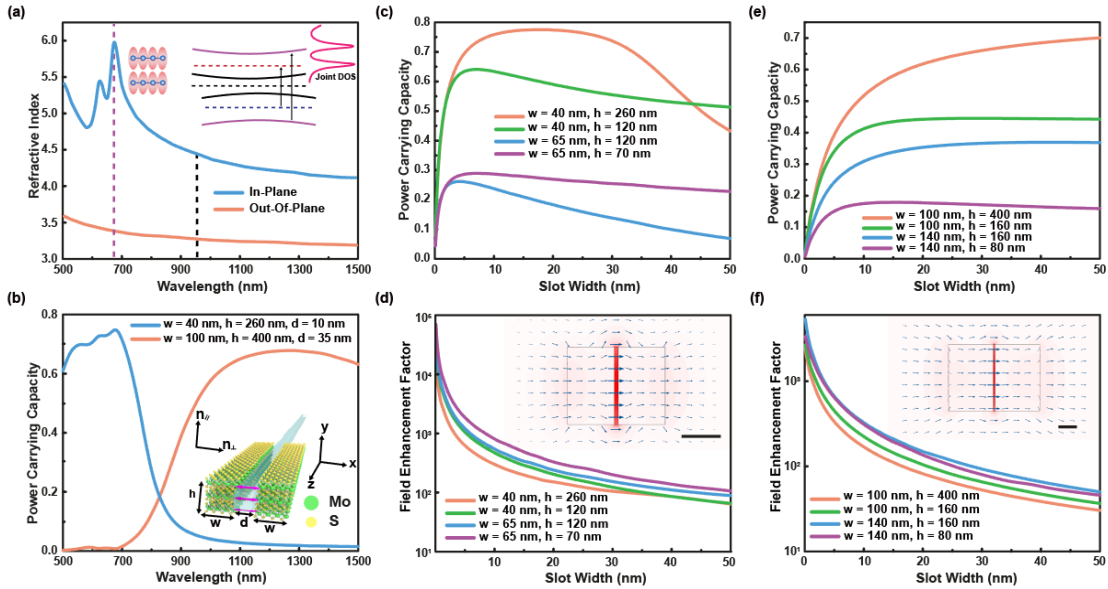

FIG. 2. Out-of-plane MoS$_2$ channel waveguide. (a) MoS$_2$ refractive indexes along in-plane and out-of-plane directions (after Ref. [67]). The purple and black dashed lines indicate the onset of MoS$_2$ excitonic resonance and bandgap optical absorption edge, respectively. Inset: Schematic illustrations of band structures and joint density of states (DOS) of bulk TMDCs which lead to high in-plane material refractive index in the visible and near-infrared ranges. (b) Calculated power carrying capacity for two

*different MoS$_2$ channel waveguides. Inset: schematic illustration of an out-of-plane channel waveguide made of bulk MoS$_2$. (c, d) Power carrying capacity and field enhancement factor for several different waveguide dimensions at $\lambda = 680\ nm$, respectively. Inset in (d): Calculated electric field and its intensity distribution at $\lambda = 680\ nm$ for a 5 nm channel. (e, f) Power carrying capacity and field enhancement factor for several different waveguide dimensions at $\lambda = 1200\ nm$, respectively. Inset in (f): Calculated electric field and its intensity distributions at $\lambda = 1200\ nm$ for a 5 nm channel. Scale bars in (d) and (f) are 50 nm.*

Another possible channel waveguide configuration – in-plane channel waveguide – is considered in Fig. 3. This geometry can be fabricated by removing individual atomic planes (Fig. 3a), as was demonstrated previously [12], providing a sub-nanometer precision control over the channel thickness. At the same time as the electric field of the fundamental guided mode is predominantly polarized along out-of-plane direction (i.e., along the crystal axis shown in Fig. 3a), the response of the waveguide is mainly driven by the lower out-of-plane component of the refractive index (Fig. 2a). The out-of-plane index is also free of exciton features leading to a lower optical loss and smaller wavelength dispersion. Importantly, an optically equivalent configuration that may be more straightforward to design and fabricate is a so-called sandwich waveguide made of a thin film of hBN inserted between MoS$_2$ multilayers (Fig. 3a) (in-plane and out-of-plane refractive index components of hBN are both ~ 2 in visible and near-infrared ranges) [70]. Such a structure possesses a similar guided mode, however due to a lower index contrast at interface (i.e., MoS$_2$-hBN versus MoS$_2$-vacuum), exhibiting weaker field confinement within the hBN channel.

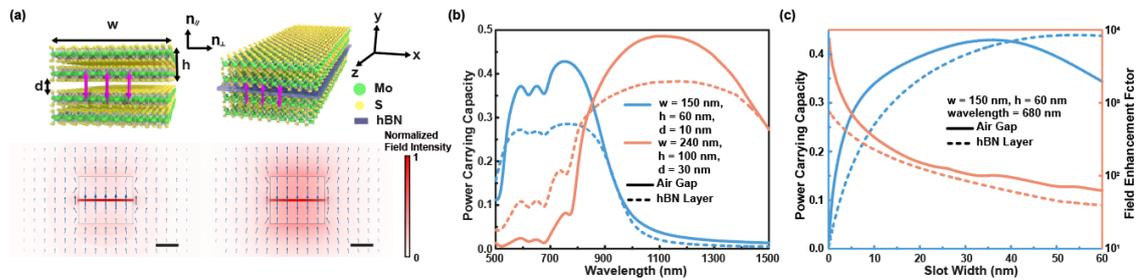

*FIG. 3. In-plane MoS$_2$ channel waveguide. (a) Schematic illustrations of in-plane channel waveguide structures: MoS$_2$-vacuum (left) and MoS$_2$-hBN (right). Respective electric field and intensity distribution at $\lambda = 1200\ nm$ are shown. Scale bar is 100 nm. (b) Calculated power carrying capacity for the two different configurations. (c) Calculated power carrying capacity and field enhancement factor for the two waveguide configurations as functions of channel width at $\lambda = 680\ nm$.*

Fig. 3b shows the dispersion of the power carrying capacity for both MoS$_2$-vacuum channel and MoS$_2$-hBN sandwich waveguides, respectively. Here again by tuning the dimensions optimal performance may be attained at a desired wavelength range. At the same time, as the out-of-plane component of

MoS$_2$ refractive index exhibits little wavelength dispersion (Fig. 2a), the variation in $C_p$ is dictated mainly by geometric dimensions. Similar to an out-of-plane channel waveguide studied in Fig. 2a, the power can be efficiently squeezed and guided in ultra-narrow channels. However, due to a smaller index contrast ($\delta n \simeq 2.4$ at MoS$_2$-vacuum interface and $\delta n \simeq 1.55$ at MoS$_2$-hBN interface), the maximum amount of optical power guided is smaller, < 50% for MoS$_2$-vacuum channel and ~35% for a respective MoS$_2$-hBN sandwich waveguide (Fig. 3b). Fig. 3c shows variation of the electric field confinement and the power carrying capacity with the channel width for $w = 150\ nm$, $h = 60\ nm$ and $\lambda = 680\ nm$. We observe that for these parameters the maximum amount of power that can be transferred in an MoS$_2$-vacuum waveguide occurs at $d = 36\ nm$ with $C_p \approx 42\%$. On the other hand, for MoS$_2$-hBN configuration exhibiting a smaller index contrast, the $C_p$ peaks at $d = 55\ nm$, with a value of 44%. Importantly, for an atomic-void configuration, i.e., when a single layer of MoS$_2$ is removed ($d = 0.7\ nm$ for in-plane channel waveguide), more than ~10% of guided power can be squeezed into the channel. On the other hand, when a single layer hBN ($d = 0.33\ nm$) is placed between such MoS$_2$ claddings, $C_p$ is only 2%. In both cases, the optical power carrying capacity is much smaller than that for an out-of-plane channel waveguide, which is expected due to a smaller index contrast at the interface. For the same reason, the maximum $C_f$ that can be achieved is about one order of magnitude (two orders of magnitude for hBN sandwich waveguides) smaller compared to that of an out-of-plane channel configuration. As the channel thickness gradually increases, $C_f$ for both configurations drops to below 100 at $d = 60\ nm$. A similar dynamics is expected in the sub-bandgap regime, e.g., at $\lambda = 1200\ nm$.

    Light propagation in such all-dielectric MoS$_2$ channel waveguides is expected to have lower optical loss when compared to plasmonic nanostructures (for example, metal-insulator-metal waveguide). However, strong excitonic response of MoS$_2$ might result in an undesired absorption [67]. To better gauge the impact of losses in our channel waveguides, we compare their performance characteristics with those of a hybrid plasmonic waveguide (Fig. 4a), which has been shown to offer an optimal tradeoff between field confinement and the optical loss [71]. We assume that the hybrid waveguide is made of a 1 nm-thick hBN film sandwiched between Au and MoS$_2$, as shown schematically in Fig. 4a. For a fair comparison, we consider a 1 nm-wide out-of-plane channel waveguide and a 1 nm-thick sandwich waveguide infiltrated with hBN. Both waveguides have similar dimensions (i.e., $d$, $w$ and $h$) as the hybrid plasmonic waveguide counterpart. The hybrid waveguide has electric field polarized along the out-of-plane direction, and thus its response is dominated by a lower loss component of the MoS$_2$ refractive index (Fig. 2a). Fig. 4b plots the wavelength dispersion of power carrying capacity and field enhancement factor for a hybrid plasmonic waveguide with $w = 200\ nm$ and $h = 80\ nm$. The guided mode is strongly confined within a 1 nm hBN layer (~3 atomic layers) and allows almost 20% of power transfer in the visible spectrum

and over 10% across the near-infrared range. The optical field in the waveguide core is also very strong; generally $C_f > 1000$ across the studied wavelength range (i.e., 500 nm – 1500 nm). The values of power carrying capacity and electric field enhancement are somewhat comparable to equivalent channel waveguide configurations studied in Figs 2 and 3 above. For example, across 750 – 1000 nm, an out-of-plane channel waveguide exhibits $C_p > 17\%$ and $C_f > 1000$. In-plane channel waveguide on the other hand exhibits lower values due to a smaller index contrast with $C_p \simeq 3.5\%$ across 750 – 1000 nm (< 1% for $\lambda > 1200\ nm$).

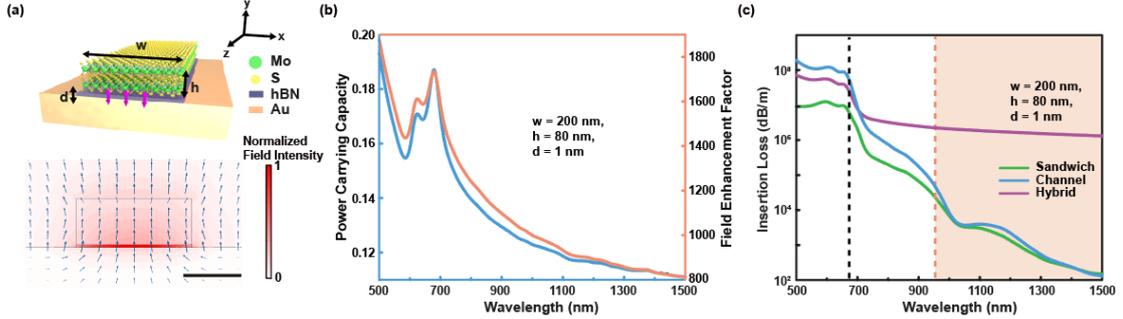

FIG. 4. Insertion loss comparison. (a) Schematic illustration of a hybrid plasmonic waveguide. Calculated electric field and its intensity distributions at $\lambda = 1200\ nm$ are also shown. Scale bar is 100 nm. (b) The wavelength dispersion of power carrying capacity and field enhancement factor of the hybrid waveguide with geometric dimensions shown in the figure. (c) The wavelength dispersion of insertion loss for a hybrid waveguide, out-of-plane and in-plane channel waveguides. The black and orange dashed lines indicate the onsets of MoS$_2$ excitonic resonance and optical bandgap absorption edge, respectively, and the shaded area represents the respective sub-bandgap regime.

Next we compare optical losses in these three waveguiding structures. Fig. 4c shows the wavelength dispersion of the insertion loss, $\gamma = \frac{20}{\ln(10)} * Im(k)$, where $k = \frac{2\pi}{\lambda}\beta$ stands for a complex wavenumber of the guided mode. Below $\lambda \simeq 680\ nm$, strong light absorption within bulk MoS$_2$ caused by excitonic resonance is the major contributor to the optical loss in all three waveguide structures. However, since for in-plane sandwich (Fig. 3a) and hybrid plasmonic (Fig. 4a) waveguides, the confined mode is polarized mainly along the crystal axis (i.e., out-of-plane direction), the associated optical loss is slightly smaller. At the same time for $\lambda > 720\ nm$, light absorption in MoS$_2$ gradually drops and reaches very small values in the sub-bandgap regime (i.e., $\lambda > 960\ nm$) as shown in Fig. 4c. Specifically, both in-plane and out-of-plane channel waveguide configurations reach about $1000\ dB/m$ at $\lambda = 1250\ nm$ and around $100\ dB/m$ at $\lambda = 1500\ nm$. As for a hybrid plasmonic waveguide, optical absorption in this wavelength range (i.e., below MoS$_2$ bandgap) is dictated by the metal substrate. As a result, the hybrid plasmonic waveguide exhibits $\gamma > 10^6\ dB/m$ in the near-infrared wavelength range. Therefore, all dielectric channel waveguides possess a much smaller insertion loss, while at the same time offering optical confinement performance comparable to that of a hybrid plasmonic waveguide.

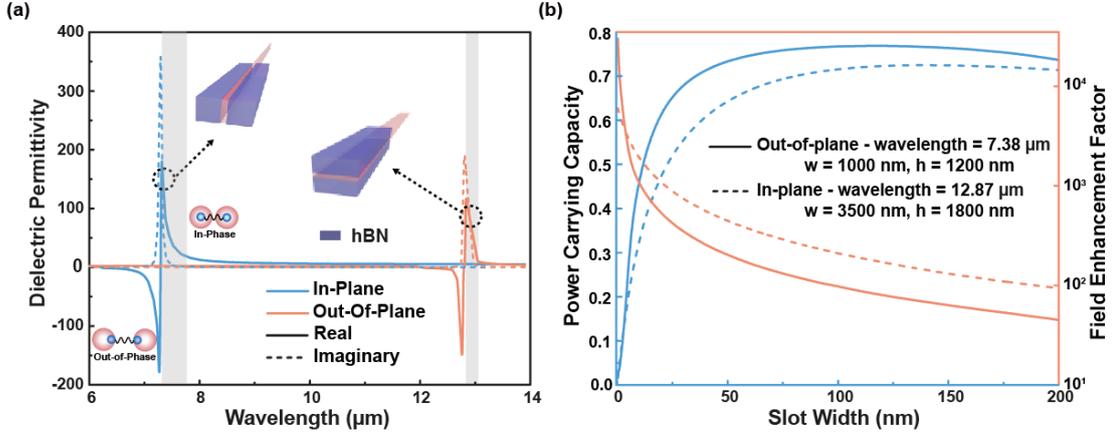

FIG. 5. hBN channel waveguides in mid-infrared range. (a) Dispersion of real and imaginary permittivities of bulk hBN along in-plane and out-of-plane orientations (after Refs. [57-60]) and schematic illustrations of in-plane and out-of-plane channel waveguides made of bulk hBN. The shaded areas indicate ranges in which large positive values of the real permittivities of bulk hBN are attained. Inset: schematic illustration of electronic and ionic dipoles when they are in phase and out of phase, respectively. (b) Calculated power carrying capacity and electric field enhancement factor for out-of-plane and in-plane hBN channel waveguides at $\lambda = 7.38\ \mu m$ and $\lambda = 12.87\ \mu m$, respectively.

Finally, high refractive index is manifested not only in excitonic TMDC materials studied above, but can also be observed in other strongly resonant material systems and at other wavelength ranges. Of a particular interest are layered VdW polaritonic materials, such as bulk hBN. hBN exhibits negative permittivity in the Reststrahlen region between transverse optical (TO) and longitudinal (LO) resonances [57-60] and this region has been studied for development of phononic equivalent of plasmonics, such as plasmonic-like waveguides (also known as interface polaritons in the condensed matter theory) [72-74] and phononic superlenses [75]. However, it is important to note that operating on the long wavelength side of the TO phonon resonance, where permittivity is positive, rather than above it, where $\varepsilon$ is negative, always leads to a superior performance [43]. This can be easily understood from the dielectric constant in the vicinity of TO resonance, following

$$\varepsilon(\omega) = \varepsilon_\infty + \frac{(\omega_{LO}^2 - \omega_{TO}^2)\varepsilon_\infty}{\omega_{TO}^2 - \omega^2 - i\omega\gamma} \qquad (2)$$

where $\gamma$ is broadening, $\varepsilon_\infty$ is high frequency (optical) dielectric constant of material due to motion of valence electrons, and the term in the numerator of the second term represents the combined strength of all the ionic oscillators. Since as mentioned above, optical response is nothing but the dipole oscillations, when one operates below the resonance, the electronic and ionic dipoles are in phase with each other and the total response is always stronger than in case of operating in the negative $\varepsilon$ region where electronic and ionic dipoles are out of phase and thus partially cancel each other (Fig. 5a inset).

As an illustration of utility of high positive and resonant permittivity, we examine channel phonon-polariton waveguides created with bulk hBN. As shown in Fig. 5a, around $\lambda = 7.3\,\mu m$, the in-plane permittivity of hBN reaches almost 200 while the out-of-plane component is only around 2 – 3. On the other hand, around $\lambda = 13\,\mu m$, the permittivity of hBN along the out-of-plane orientation is about 100 while the in-plane component is about 5. As such we expect that out-of-plane channel waveguide configuration would be suitable for $\lambda \simeq 7.3\,\mu m$, whereas the in-plane channel waveguide configuration for $\lambda \simeq 13\,\mu m$. In Fig. 5b we plot respective power carrying capacity and field enhancement factor as functions of the channel width, $d$, for in-plane and out-of-plane waveguide configurations at respective wavelengths. Both mid-infrared hBN channel waveguides can transfer over 70% of the power within only 150 nm wide channel, which is less than 2% of the free-space wavelength. Moreover, mid-infrared light with several microns of free-space wavelength can be efficiently confined within sub-nanometer channels and with very strong electric field enhancement factor (20000 for the out-of-plane configuration and over 5000 for the in-plane one).

**Conclusion**

Our study shows that light can be efficiently confined and guided in ultranarrow VdW channels with atomic size features. Extremely deep subwavelength confinement (<λ/100) and strong electric field enchantment are of great promise for enhancing light-materials interaction at nanometer and sub-nanometer scales. We foresee that our findings will find use in high precision optical sensing, nanoscale opto-mechanics, and high efficiency, small footprint interconnects, among others.

**Methods**

Numerical Simulations: Mode dispersion and power and electric field distribution of the one dimensional channel waveguides are calculated with MATLAB. Mode analysis simulations of 2D channel waveguides and hybrid plasmonic waveguides are performed with a finite element method. All the simulations are performed with no substrates and background material being air.

**Acknowledgement**
A.R.D. acknowledges support by UCLA Faculty Research Grant, the Hellman Society of Fellows, and partial support by NASA Innovative Advanced Studies

(grant 80NSSC21K0954). J.B.K. acknowledges support by DARPA Award # HR00111720032## References

(1) Novoselov, K. S.; Jiang, D.; Schedin, F.; Booth, T. J.; Khotkevich, V. V.; Morozov, S. V.; Geim, A. K. Two-Dimensional Atomic Crystals. *PNAS* **2005**, 102, 10451-10453.

(2) Wang, Q. H.; Kalantar-Zadeh, K.; Kis, A.; Coleman, J. N.; Strano, M. S. Electronics and Optoelectronics of Two-Dimensional Transition Metal Dichalcogenides. *Nat. Nanotechnol.* **2012**, 7, 699-712.

(3) Geim, A. K.; Grigorieva, I. V. Van der Waals Heterostructures. *Nature* **2013**, 499, 419-425.

(4) Das, S.; Robinson, J. A.; Dubey, M.; Terrones, H.; Terrones, M. Beyond Graphene: Progress in Novel Two-Dimensional Materials and van der Waals Solids. *Annu. Rev. Mater. Res.* **2015**, 45, 1-27.

(5) Novoselov, K. S.; Mishchenko, O. A.; Carvalho, O. A.; Castro Neto, A. H. 2D Materials and van der Waals Heterostructures. *Science* **2016**, 353, aac9439.

(6) Mak, K. F.; Shan, J. Photonics and Optoelectronics of 2D Semiconductor Transition Metal Dichalcogenides. *Nat. Photonics* **2016**, 10, 216-226.

(7) Manzeli, S.; Ovchinnikov, D.; Pasquier, D.; Yazyev, O. V.; Kis, A. 2D Transition Metal Dichalcogenides. *Nat. Rev. Mater.* **2017**, 2, 1-15.

(8) Blees, M.K.; Barnard, A.W.; Rose, P.A.; Roberts, S.P.; McGill, K.L.; Huang, P.Y.; Ruyack, A.R.; Kevek, J.W.; Kobrin, B.; Muller, D.A; McEuen, P.L. Graphene Kirigami. *Nature* **2015**, 524, 204-207.

(9) Geim, A. K. Exploring Two-Dimensional Empty Space. *Nano Lett.* **2021**, 21, 6356-6358.

(10) Munkhbat, B.; Canales, A.; Küçüköz, B.; Baranov, D. G.; Shegai, T. O. Tunable Self-Assembled Casimir Microcavities and Polaritons. *Nature* **2021**, 597, 214-219.

(11) Radha, B.; Esfandiar, A.; Wang, F.C.; Rooney, A.P.; Gopinadhan, K.; Keerthi, A.; Mishchenko, A.; Janardanan, A.; Blake, P.; Fumagalli, L.; Lozada-Hidalgo, M. Molecular Transport through Capillaries Made with Atomic-Scale Precision. *Nature* **2016**, 538, 222-225.

(12) Keerthi, A.; Geim, A.K.; Janardanan, A.; Rooney, A.P.; Esfandiar, A.; Hu, S.; Dar, S.A.; Grigorieva, I.V.; Haigh, S.J.; Wang, F.C.; Radha, B. Ballistic Molecular Transport through Two-Dimensional Channels. *Nature* **2018**, 558, 420-424.

(13) Fumagalli, L.; Esfandiar, A.; Fabregas, R.; Hu, S.; Ares, P.; Janardanan, A.; Yang, Q.; Radha, B.; Taniguchi, T.; Watanabe, K.; Gomila, G. Anomalously Low Dielectric Constant of Confined Water. *Science* **2018**, 360, 1339-1342.

(14) Mouterde, T.; Keerthi, A.; Poggioli, A.R.; Dar, S.A.; Siria, A.; Geim, A.K.; Bocquet, L.; Radha, B. Molecular Streaming and Its Voltage Control in Ångström-Scale Channels. *Nature* **2019**, 567, 87-90.

(15) Kim, K.; Song, B.; Fernández-Hurtado, V.; Lee, W.; Jeong, W.; Cui, L.;